\documentclass[letterpaper,11pt]{article}
\usepackage{caption}
\pdfoutput=1 

\usepackage{extarrows}
\usepackage{amssymb}
\usepackage{amsmath}

\usepackage{amstext}
\usepackage{etoolbox}
\let\tinymatrix\smallmatrix

\patchcmd{\tinymatrix}{\scriptstyle}{\scriptscriptstyle}{}{}
\patchcmd{\tinymatrix}{\scriptstyle}{\scriptscriptstyle}{}{}
\patchcmd{\tinymatrix}{\vcenter}{\vtop}{}{}
\patchcmd{\tinymatrix}{\bgroup}{\bgroup\scriptsize}{}{}

\usepackage{graphicx,epsfig}
\usepackage{epsfig}
\usepackage{verbatim} 
\usepackage{color}
\usepackage{ulem}
\usepackage{enumitem}
\usepackage{subfigure}
\usepackage{bbm}
\usepackage{parskip}
\usepackage{dsfont}
\usepackage[numbers,sort&compress]{natbib}
\usepackage[body={6.5in, 9in}, right=1in, top=1in]{geometry}
\usepackage{mathtools}
\usepackage{comment}
\usepackage{caption}
\usepackage{bbold}
\usepackage{float}

\newcommand{\Comment}[1]{{}}
\definecolor{darkblue}{rgb}{0.15,0.35,0.55}
\definecolor{reddish}{rgb}{0.65, 0.2, 0.2}
\usepackage[linktocpage=true]{hyperref}

\newcommand{\be}{\begin{equation}}
\newcommand{\ee}{\end{equation}}
\newcommand{\bea}{\begin{eqnarray}}
\newcommand{\eea}{\end{eqnarray}}
\newcommand{\beas}{\begin{eqnarray*}}
\newcommand{\eeas}{\end{eqnarray*}}

\def\({\left(}
\def\){\right)}

\def\gsim{ \lower .75ex \hbox{$\sim$} \llap{\raise .27ex \hbox{$>$}} }
\def\lsim{ \lower .75ex \hbox{$\sim$} \llap{\raise .27ex \hbox{$<$}} }

\usepackage{xcolor,colortbl}

\begin{document}
\def\thefootnote{\fnsymbol{footnote}}

\begin{center}
\LARGE{\textbf{The Self-Organized Critical Multiverse}} \\[0.5cm]
 
\large{Guram Kartvelishvili, Justin Khoury, and Anushrut Sharma}
\\[0.5cm]

\small{
\textit{Center for Particle Cosmology, Department of Physics and Astronomy, University of Pennsylvania,\\ Philadelphia, PA 19104}}

\vspace{.2cm}

\end{center}

\vspace{.6cm}

\hrule \vspace{0.2cm}
\centerline{\small{\bf Abstract}}
{\small\noindent 
Recently a dynamical selection mechanism for vacua based on search optimization was proposed in the context of false-vacuum eternal inflation on the landscape. 
The search algorithm, defined by local vacuum transitions, is optimal in regions of the landscape where the dynamics are tuned at criticality, 
with de Sitter vacua having an average lifetime of order their Page time. The purpose of this paper is to shed light on the nature of the dynamical phase transition at the Page lifetime.
We focus on a finite region of the landscape, which exchanges volume with the rest of the landscape and as such acts as an open system. Through
a change of variables the master equation governing the comoving volume of de Sitter vacua is mapped to a stochastic equation for coupled overdamped
stochastic oscillators --- the well-known Ornstein-Uhlenbeck process. The rest of the landscape, which acts as an environment, is assumed to result in a non-vanishing driving term for all sites in the region with uncorrelated, white noise fluctuations (though not necessarily Gaussian). We first show that the static susceptibility of the oscillators diverges as the average lifetime of de Sitter vacua approaches the Page time. Thus, optimal regions of the landscape are most susceptible to volume influx from their environing landscape. We then show that the displacement fluctuations for the oscillators exhibit a~$1/f$ power spectrum over a broad range of frequencies, precisely at the critical Page lifetime distribution. A~$1/f$ power spectrum is a hallmark of non-equilibrium systems at criticality. In analogy with sand avalanches in the abelian sandpile or neuronal avalanches in the brain, de Sitter vacua at criticality can be thought of as undergoing scale invariant
volume fluctuation avalanches.  
\vspace{0.3cm}
\noindent
\hrule
\def\thefootnote{\arabic{footnote}}
\setcounter{footnote}{0}

\section{Introduction}

The discovery that string theory admits a vast landscape of metastable vacua~\cite{Bousso:2000xa,Kachru:2003aw}, together with the mechanism of eternal inflation~\cite{Starobinsky:1982ee,Vilenkin:1983xq,Linde:1986fc,Linde:1986fd,Starobinsky:1986fx} for dynamically populating these vacua, has led to a paradigm shift in our understanding of fundamental physics. It entails that statistical physics, possibly in conjunction with selection (anthropic) effects, played a role in determining the physical parameters of our universe. Like many other statistical systems, it is natural to expect that the multiverse can exhibit phase transitions. Indeed, it has been shown recently that certain regions of the landscape display non-equilibrium critical phenomena, in the sense that their vacuum dynamics are tuned at {\it dynamical criticality}~\cite{Khoury:2019yoo,Khoury:2019ajl}. 

Unlike standard approaches to the measure problem, {\it e.g.},~\cite{Garriga:1997ef,Garriga:2005av}, which focus on the asymptotic quasi-stationary probability distribution of vacua, the analysis of~\cite{Khoury:2019yoo,Khoury:2019ajl} focused on the {\it approach to equilibrium}. This is motivated by the possibility that eternal inflation has unfolded for a time much shorter than the exponentially long relaxation time for the landscape.\footnote{Since in eternal inflation the vast majority of observers are produced at asymptotically late times, the ``early-time" assumption amounts to an assumption that we are atypical. As emphasized by Hartle and Srednicki~\cite{Hartle:2007zv,Srednicki:2009vb}, in a situation like eternal inflation where our current data is replicated at (infinitely-many) other space-time locations, one must make an assumption about our location within the multiverse to extract any physical predictions. This can be formalized in a Bayesian sense by introducing what Hartle and Srednicki call a xerographic probability distribution~\cite{Hartle:2007zv,Srednicki:2009vb}. The ``early-time" assumption amounts to a xerographic distribution which cuts off at times much earlier than the relaxation time. This is in contrast with the typicality (or principle of mediocrity) assumption, which amounts to a uniform xerographic distribution. A key point of~\cite{Hartle:2007zv,Srednicki:2009vb} is that the choice of xerographic distribution is observationally testable, for instance by comparing likelihoods of our current data or making predictions for future measurements. We will make this more precise in a forthcoming paper~\cite{to appear Sam}.} In this case, as first emphasized in~\cite{Denef:2017cxt}, a vacuum like ours should be likely not because it is typical according to the stationary distribution, but rather because it had the right properties to be accessed early on in the evolution. This perspective offers a dynamical selection mechanism for vacua based on search optimization~\cite{Khoury:2019yoo}: vacua that are easily accessed reside in optimal regions where the search algorithm defined by local landscape dynamics is efficient. This idea was formalized with the definition of an accessibility measure~\cite{Khoury:2019ajl}, which is the landscape analogue of the closeness centrality index~\cite{closeness1,closeness2} in network theory.

At times much smaller than the mixing time, most hospitable vacua have been accessed once or perhaps not at all. The statistics best-suited for such
a situation are first-passage statistics, which capture the probability for vacua to be visited for the first time. In~\cite{Khoury:2019yoo}, search optimization was defined
by two competing requirements:

\begin{itemize} 

\item Search efficiency, which requires minimizing the mean first-passage time (MFPT) to vacua. It was shown that the MFPT is minimized
for vacua lying at the bottom of funnel-like regions of the landscape, akin to the smooth folding funnels of naturally-occurring proteins~\cite{proteins}. 

\item Sweeping exploration, which is enforced by demanding that the Markov process describing landscape dynamics is recurrent, 
even in the limit of infinite region size. In recurrent random walks every vacuum is guaranteed to be visited. 

\end{itemize}

\noindent Optimal regions of the landscape reach a compromise between minimal oversampling and sweeping exploration. 
This is achieved by having the smallest MFPT compatible with recurrence. Thus, optimal regions lie at the critical boundary
between recurrence and transience, with the average MFPT scaling logarithmically with the number of vacua, which signals
dynamical criticality. This is analogous to the non-equilibrium phase transition in quenched disordered media, when the probability
distribution of waiting times reaches a critical power-law~\cite{disordered media}. 

A key result of~\cite{Khoury:2019yoo,Khoury:2019ajl} is that dynamical criticality on the landscape corresponds to de Sitter (dS) vacua having an average lifetime given by
the dS Page time~\cite{Page:1993wv}:~$\tau_{\rm crit} \sim M_{\rm Pl}^2/H^3$.\footnote{In the context of black holes, the Page time corresponds to the half-evaporation time~\cite{Page:1993wv} and marks the transition when the entanglement entropy between the Hawking radiation and the black hole begins to decrease. The Page time in dS is analogously defined by replacing the Schwarzchild radius with the Hubble radius. It is motivated by the well-known similarities in causal and thermodynamical properties of black hole and de Sitter geometries, {\it e.g.},~\cite{Danielsson:2002td,Danielsson:2003wb,Ferreira:2016hee,Ferreira:2017ogo}.} In the context of slow-roll inflation, the Page time corresponds to the time when the curvature perturbation reaches unit amplitude, which marks the phase transition to slow-roll eternal inflation~\cite{Creminelli:2008es,Dubovsky:2008rf,Dubovsky:2011uy}. It has also been used to place a bound on the maximum number of e-folds that can be described semi-classically~\cite{ArkaniHamed:2007ky}. In the case of pure dS, the coincident graviton 2-point function~\cite{Starobinsky:1979ty}, like that of any massless field in dS ({\it e.g.},~\cite{Linde:2005ht}), grows linearly in time,~$\langle h^2(x) \rangle \sim H^3 \tau /M_{\rm Pl}^2$, and becomes~${\cal O}(1)$ around the Page time, signaling a breakdown of perturbation theory.\footnote{We thank an anonymous referee for pointing this out to us.}  The appearance of the Page time in the present context of false-vacuum eternal inflation, and its relation to dynamical criticality, is surprising and deserves a deeper understanding. In any case, for our vacuum the predicted lifetime is~$M_{\rm Pl}^2/H_0^3 \sim 10^{130}~{\rm years}$, which remarkably agrees to within~$\sim 2\sigma$ with the Standard Model estimate for Higgs metastability~\cite{EliasMiro:2011aa,Andreassen:2017rzq}.

The goal of this paper is to gain further insights on the nature of the dynamical phase transition at the Page lifetime. For this purpose, we consider
a finite region of the landscape comprised of~$N\gg 1$ dS vacua, as well as terminal (AdS or Minkowski) vacua. Vacua in the region interact
({\it i.e.}, exchange volume) with the rest of the landscape, which we treat as an environment. Thus, the master equation governing the volume of dS vacua includes a stochastic source term from the environment. Since we are ultimately interested in the region of interest being nearly optimal, there is a natural separation of time scales: transition rates between dS vacua are rapid compared to the slow evolution of the environment. Therefore, the region of interest is {\it open, slowly-driven and out-of-equilibrium.}

Through a change of variables, we will show that the master equation can be mapped to the well-known
Ornstein-Uhlenbeck process~\cite{handbook} describing a system of~$N$ coupled overdamped stochastic oscillators. This
allows us to understand criticality at the Page lifetime as a dynamical phase transition in the Ornstein-Uhlenbeck process.

We first study in Sec.~\ref{chi} the susceptibility of the oscillators, describing their response to external driving.
We will show that the mean static susceptibility exhibits a divergence as the average proper lifetime of vacua approaches
the Page time. A divergent susceptibility is one of the hallmarks of systems at criticality. This offers an alternative interpretation of
the optimal compromise offered by the Page lifetime: regions of dS vacua with such lifetimes are most susceptible to volume influx
from their environment, while at the same time minimizing their MFPT.

In Sec.~\ref{1/f} we turn to another signature of non-equilibrium systems at criticality: a~$1/f$ (or ``pink" noise) fluctuation power spectrum. 
We will show that the displacement fluctuations for the oscillators exhibit a~$1/f$ signal precisely at the critical Page lifetime
distribution. The origin of the~$1/f$ spectrum can be understood intuitively from the linearity of the Ornstein-Uhlenbeck process.
The auto-correlation function for fluctuations is a weighted linear superposition of processes with different characteristic time scales,
set by the decay rates of vacua in the region. The Page lifetime corresponds to a uniform distribution of decay rates, resulting in a
superposition with~$1/f$ spectrum. This is analogous to an old argument for the origin of flicker noise as an aggregation of shot noise processes~\cite{vanderziegle}.
(See also~\cite{origin1overfMarkov,origin1overf}.)

Non-equilibrium systems exhibiting~$1/f$ fluctuation spectra are ubiquitous in nature. Examples include neuronal dynamics~\cite{brain}, heart beat variability~\cite{heart},
linguistics (Zipf's law), economic time series~\cite{econ} ({\it e.g.}, stock market prices),
music~\cite{Voss}\footnote{Evidently we are drawn to music with~$1/f$ spectrum because it is neither too random (like white noise) nor too
predictable (like Brownian~$1/f^2$ noise).} and art~\cite{paintings} ({\it e.g.}, Pollock's paintings~\cite{Pollock}). Thus, complex behavior appears intimately related to dynamical criticality. 
This has motivated the tantalizing idea of self-organizing criticality~\cite{SOC1,SOC2,SOCbook}. While the subject is not without controversy,
it is worth noting that our framework satisfies what are believed to be necessary conditions for self-organized criticality --- our landscape region is
out-of-equilibrium, open/dissipative, and slowly-driven. In analogy with sand avalanches in the abelian sandpile~\cite{SOC1}, or neuronal avalanches in the brain, 
dS vacua at criticality can be thought of as undergoing scale invariant volume fluctuation avalanches.

\section{Master Equation for Landscape Dynamics}
\label{RW complex nets}

Consider a finite region of the landscape comprised of~$N \gg 1$ inflating vacua and a certain number of terminal (AdS or Minkowski) vacua. The region is surrounded by a much larger
landscape of inflating and terminal vacua. In what follows we use indices~$i,j = 1,\ldots, N$ to denote dS vacua inside the region, indices~$\alpha,\beta,\ldots$ for dS vacua outside the region, and indices~$a,b,\ldots$ to collectively denote all terminal vacua (both inside and outside). 

Let~$f_i(t)$ denote the fraction of total comoving volume occupied by dS vacuum~$i$ in the region, as a function of the global time variable~$t$  parametrizing the foliation. 
Upon coarse-graining over a time interval longer than transients between periods of vacuum energy domination, the~$f_i$'s satisfy the master equation~\cite{Garriga:1997ef,Garriga:2005av}:
\be
\frac{{\rm d}f_i}{{\rm d}t}  = \sum_j M_{ij} f_j  +  \sum_\alpha \kappa_{i\alpha}f_\alpha \,.
\label{master 1}
\ee
Here~$M_{ij}$ is the transition matrix, 
\be
M_{ij} \equiv \kappa_{ij} - \delta_{ij} \kappa_j \,,
\label{M def}
\ee
where~$\kappa_{ij}$ is the~$j \rightarrow i$ transition rate, and~$\kappa_i$ is the total decay rate of vacuum~$i$:
\be
\kappa_i \equiv \sum_j \kappa_{ji} + \sum_\alpha \kappa_{\alpha i} + \sum_a \kappa_{ai} \,.
\ee
This encodes the fact that~$i$ can decay respectively into other dS vacua in the region, to dS vacua outside the region, and to terminal 
vacua both inside and outside the region.

For concreteness transitions are assumed to be governed by Coleman-De Luccia (CDL) instantons~\cite{Coleman:1977py,Coleman:1980aw}.
For~${\rm dS}\rightarrow {\rm dS}$ transitions, the CDL rate is of the form
\be
\kappa_{ij} = \frac{A_{ij}}{w_j} \,.
\label{kappa dSdS}
\ee
Here~$A_{ij} = A_{ji}$ is a symmetric adjacency matrix~\cite{Lee:1987qc} given by~$A_{ij} = \left(\Lambda^4 {\rm e}^{-S_{\rm bounce}}\right)_{ij}$, where~$S_{\rm bounce}$ is the Euclidean action of the bounce solution and~$\Lambda^4$ is the fluctuation determinant. Meanwhile,~$w_j$ is the `weight':
\be
w_j = H_j^3 {\cal N}_j^{-1} {\rm e}^{S_j}\,,
\label{weights}
\ee
with~$S_j$ denoting the dS entropy of the parent vacuum. The factor of~$H_j^3 = \left(V_j/3M_{\rm Pl}^2\right)^{3/2}$ converts the CDL rate per unit volume to a transition rate,
while the lapse function~${\cal N}_j^{-1}$ converts the rate from unit proper time to global time via~$\Delta \tau_j = {\cal N}_j \Delta t$. We will not need the explicit form of the~${\rm dS}\rightarrow {\rm AdS/Minkowski}$ transition rate, but suffice to say that it takes a similar form to~\eqref{kappa dSdS} except that the numerator is not symmetric. 

Although~$M_{ij}$ is not symmetric, it nevertheless has real eigenvalues, and its eigenvectors form a complete basis of the~$N$-dimensional space of dS vacua in the region.
This can be seen by performing a similarity transformation,
\be
\Sigma = W^{-1/2} M \, W^{1/2}\,;\qquad  W \equiv {\rm diag}(w_1,w_2,\ldots,w_{N})\,. 
\label{Sig M}
\ee
Using~\eqref{kappa dSdS} it is easy to see that~$\Sigma$ is symmetric and therefore has real eigenvalues. Moreover, since~\eqref{Sig M} defines a similarity transformation,~$\Sigma$
and~$M$ have identical spectra. Assuming that~$M$ is irreducible, {\it i.e.}, there exists a sequence of transitions connecting any pair of inflating vacua in the region, 
it follows from Perron-Frobenius' theorem that its largest eigenvalue is non-degenerate and negative,~$\lambda_1 < 0$, while all other eigenvalues are strictly smaller:
\be
0 > \lambda_1 > \lambda_2 \geq \ldots \geq \lambda_{N}\,.
\label{lamba's M}
\ee
Thus,~$\Sigma$ is negative definite. The eigenvectors of~$\Sigma$, denoted by~$v^{(\ell)}$,~$\ell = 1,\ldots, N$, form a complete and orthonormal basis for dS vacua in the region:
$\sum_{\ell =1}^{N} v^{(\ell)}_i v^{(\ell)}_j  = \delta_{ij}$,~$\sum_i v^{(\ell)}_i v^{(\ell')}_i = \delta^{\ell \ell'}$. The eigenvectors of~$M$ are obtained by a simple rescaling, 
$v^{(\ell)}_M = W^{1/2}v^{(\ell)}$.

Coming back to the~${\rm dS}\rightarrow {\rm dS}$ transition rates~\eqref{kappa dSdS}, it follows from the symmetry of the adjacency matrix,~$A_{ij} = A_{ji}$,
that such rates satisfy detailed balance:
\be
\frac{\kappa_{ji}}{\kappa_{ij}} = \frac{w_j}{w_i}\sim {\rm e}^{S_j-S_i}\,.
\label{detailed balance}
\ee
An immediate consequence is that upward tunneling is exponentially suppressed compared to downward tunneling. This allows one to define a ``downward" approximation in which upward tunneling is neglected to leading order~\cite{SchwartzPerlov:2006hi,Olum:2007yk}. By labeling dS vacua in order of increasing potential energy,~$0 < V_1 \leq \ldots \leq V_{N}$, the transition matrix becomes upper-triangular in this approximation, with diagonal entries~$-\kappa_j$. Thus, the eigenvalues are simply given by the decay rates of vacua, 
\be
\{\lambda_1,\ldots,\lambda_N\} \simeq {\rm sort}\{-\kappa_1,\ldots,-\kappa_N\}\,,
\label{downward}
\ee
where ``sort" stands for ordering the~$\kappa$'s from smallest to largest. In particular,~$\lambda_1$ is set by the slowest decay rate, while
the other eigenvalues~$\lambda_2,\ldots, \lambda_{N}$ are given by the decay rates of dS vacua in increasing order of instability.

\section{Mapping to Ornstein-Uhlenbeck Process}
\label{map OU}

The master equation~\eqref{master 1} for dS vacua in the region can be mapped to~$N$ stochastic overdamped coupled oscillators,
corresponding to the well-known Ornstein-Uhlenbeck process~\cite{handbook}. This will allow us to understand criticality at the Page lifetime
as a dynamical phase transition in the Ornstein-Uhlenbeck process.

To see this, first rewrite~\eqref{master 1} in terms of the comoving volume~${\cal V}_i$ of each dS vacuum in the region. Since the total comoving volume is conserved, the translation is obvious:
\be
\frac{{\rm d}{\cal V}_i}{{\rm d}t}  = \sum_j M_{ij} {\cal V}_j  +  \sum_\alpha \kappa_{i\alpha}{\cal V}_\alpha  \,.
\label{master 2}
\ee
By changing variables to
\be
Q = W^{-1/2} {\cal V}\,; \qquad K_{ij} = -\Sigma_{ij}\,,
\ee
the master equation can be recast as
\be
\frac{{\rm d}Q_i}{{\rm d}t}  = -\sum_j K_{ij} Q_j  +  \sum_\alpha \sqrt{\frac{w_\alpha}{w_i}}\kappa_{i\alpha} Q_\alpha  \,.
\label{osc 1}
\ee
Note that $K_{ij}$ is symmetric and positive definite, with eigenvalues $-\lambda_\ell$. 
Equation~\eqref{osc 1} describes~$N$ overdamped oscillators, coupled to a reservoir of other oscillators. 

From the point of view of the region of interest, the last term in~\eqref{osc 1} amounts to a driving force. In matrix form,
\be
\frac{{\rm d}Q}{{\rm d}t} = - K Q + \eta\,;\qquad \eta_i \equiv  \sum_\alpha \sqrt{\frac{w_\alpha}{w_i}}\kappa_{i\alpha} Q_\alpha\,.
\label{osc eta}
\ee
The driving term is of course stochastic, owing to the time-dependence of the environmental oscillators. 
Importantly, as justified in the Appendix, the time evolution of~$\eta$ is oblivious to perturbations from the region itself, as required
for the evolution of the region to be approximately Markovian. Specifically, we imagine that the environment consists of two distinct sets
of vacua: a set of high energy dS vacua, which source the region, and a set of much lower energy vacua (including terminals), which act as sinks. 
This distinction is justified by the downward approximation. Thus no vacuum in the environment plays a dual role of receiving volume from some vacuum~$j$ in the region
and giving it back to another vacuum~$i$ in the region.

Since ultimately we will be interested in optimal regions of the landscape with rapid evolution, we can treat the environment
as slowly-evolving compared to the oscillators of interest. In other words, we assume that the driving is slow compared to
the eigenvalues of~$K$. In this case we can decompose~$\eta$ into a nearly constant piece and a noise term:
\be
\eta = \bar{\eta} + \xi(t)\,.
\label{eta fluc}
\ee
For simplicity, in what follows we will assume that~$\eta_i\neq 0$ for all~$i$, that is, every oscillator in the
region is coupled to at least one oscillator in the environment. While not strictly necessary, this 
assumption will simplify the analysis below. One can think of~$\bar{\eta}$ as the average of the driving term over the (relatively short) relaxation time within the region. 
This constant piece sets the non-equilibrium steady state position of the oscillators:
\be
\bar{Q} = K^{-1} \bar{\eta}\,.
\label{osc eq}
\ee
The deviation from steady state,~$q = Q- \bar{Q}$, satisfies
\be
\frac{{\rm d}q}{{\rm d}t}   = - K q + \xi(t)\,.
\label{osc 2}
\ee
For simplicity we model the noise as uncorrelated and white:
\be
\langle \xi_i(t) \rangle = 0\,;\qquad \langle \xi_i(t)\xi_j(t') \rangle = \frac{2 D}{\Delta t} \delta_{ij} \delta(t-t')\,,
\label{noise}
\ee
where the brackets denote as usual the ensemble average, and the factor of~$\Delta t$ is introduced for convenience.
In what follows we will only need the first two moments of $\xi$, specified by~\eqref{noise}. Importantly
we need not impose that the noise is Gaussian. Equations~\eqref{osc 2} and~\eqref{noise} define an Ornstein-Uhlenbeck with uncorrelated white noise. We leave to
future work a study of more general environmental driving scenarios, such as non-vanishing driving for only a subset of sites, as
well as more general noise models. In this paper we proceed with what is arguably the simplest possibility, namely a non-vanishing
driving term for all sites ($\eta_i\neq 0$ for all $i$) with uncorrelated white noise fluctuations.

\section{Divergent Susceptibility at the Page Lifetime}
\label{chi}

In this Section we will show that the mean static susceptibility of the oscillators begins to diverge precisely when the average lifetime of dS vacua in the region approaches their Page time from below.
A divergent susceptibility is a hallmark of systems at criticality.  

To see this, we treat~$\eta$ in~\eqref{osc eta} as a non-fluctuating source and study the response of the oscillators. The solution follows readily by Fourier transform:
\be
Q(\omega) = \frac{1}{-{\rm i}\, \omega\mathbb{1} + K}\, \eta(\omega)\,.
\ee
From this we can immediately read off the frequency-dependent {\it susceptibility} tensor,
\be
\chi(\omega) =  \frac{1}{\left(-{\rm i}\, \omega\mathbb{1} + K\right)\Delta t}\,,
\ee
where the factor of~$\Delta t$ has been included to make~$\chi(\omega)$ dimensionless and, as we will see, time-reparametrization invariant. 
The average susceptibility is obtained by taking the trace, 
\be
\chi_{\rm avg}(\omega) \equiv \frac{1}{N} {\rm Tr}\,\chi(\omega) = -\frac{1}{N\Delta t} \sum_{\ell = 1}^N  \frac{1}{{\rm i}\, \omega + \lambda_\ell}\,.
\ee

This expression greatly simplifies in the ``downward" approximation~\eqref{downward}, where the eigenvalues are given by the decay rates of the dS vacua:
\be
\chi_{\rm avg}(\omega) \simeq  \frac{1}{N} \sum_{j = 1}^N  \frac{1}{-{\rm i}\, \hat{\omega} + \hat{\kappa}_j }\qquad (\text{downward})\,.
\ee
Here~$\hat{\kappa}_j \equiv \kappa_j \Delta t$ is a dimensionless (and time-reparametrization invariant) transition probability, and~$\hat{\omega} = \omega \Delta t$ is a dimensionless frequency.
Clearly the susceptibility is time-reparametrization invariant, as claimed earlier. Since~$N\gg 1$ the sum can be approximated as an integral,
\be
\chi_{\rm avg}(\omega) \simeq \int_{\hat{\kappa}_{\rm min}}^{\hat{\kappa}_{\rm max}} {\rm d}\hat{\kappa}\, \frac{\rho(\hat{\kappa})}{-{\rm i}\, \hat{\omega} + \hat{\kappa}} \,,
\label{chi tot kappa}
\ee
where~$\rho(\hat{\kappa})$ is the underlying probability distribution of such rates. The limits of integration,~$\hat{\kappa}_{\rm min}$ and~$\hat{\kappa}_{\rm max}$, are respectively the smallest and largest decay rates achieved in the region. 

As discussed in~\cite{Khoury:2019yoo,Khoury:2019ajl}, the decay rate~$\kappa_i$ of a given dS vacuum in general depends both on its potential energy~$V_i$ as well as the shape of the surrounding potential barriers. Assuming for simplicity that the absolute height of a vacuum and the shape of potential barriers are uncorrelated, we can marginalize over ``barrier parameters" to obtain an average 
decay rate~$\kappa (V)$ for vacua of given potential energy~$V$. Assuming as usual that the probability distribution of potential energy~$\rho(V)$ is nearly uniform for~$V$ much smaller than the fundamental scale~\cite{Weinberg:2000qm},~\eqref{chi tot kappa} becomes
\be
\chi_{\rm avg}(\omega) \simeq \int_{V_{\rm min}}^{V_{\rm max}} \frac{{\rm d}V}{-{\rm i}\, \hat{\omega} + \hat{\kappa}(V)} \,.
\label{chi tot V}
\ee

We will be primarily interested in the lower end of the integral, around~$V_{\rm min}$. For a uniform distribution, the smallest potential energy is on average inversely proportional to the number of vacua,~$V_{\rm min} \sim M_{\rm Pl}^4/N$. Furthermore, given our assumption that the environment is slowly-driving the region, we can focus on the zero-frequency limit, or static susceptibility~$\chi_{\rm avg}(0)$. Clearly the static susceptibility will converge or diverge as~$V_{\rm min}\rightarrow 0$ depending on whether~$\hat{\kappa}(V)$
goes to zero slower or faster than linearly in~$V$. In other words, assuming for simplicity a power-law form~$\hat{\kappa}(V)\sim V^{1+\alpha}$ as~$V\rightarrow 0$, which corresponds to~$\rho(\hat{\kappa}) \sim \hat{\kappa}^{-\frac{\alpha}{1+\alpha}}$, it is easy to see that the static susceptibility~$\chi_{\rm avg}(0)$ diverges for any $\alpha \geq 0$. The critical distribution, corresponding to a logarithmic divergence, is obtained for $\alpha \rightarrow 0$,
\be
\hat{\kappa}_{\rm crit}(V) \sim \frac{V}{M_{\rm Pl}^4}\sim \frac{H^2}{M_{\rm Pl}^2}\,,
\label{kap crit 1}
\ee
where we have used~$M_{\rm Pl}$, the natural scale in the problem, to fix dimensions. Since the probability distribution of potential energy~$\rho(V)$ was assumed nearly uniform, the critical case corresponds to a flat distribution of decay rates:
\be
\rho_{\rm crit}(\hat{\kappa}) = {\rm const}\,.
\label{rho crit}
\ee

It is straightforward to translate~\eqref{kap crit 1} to a critical decay rate in units of proper time using~$\hat{\kappa} = \kappa_i^{\rm proper} \Delta \tau$. The natural proper time step is of course the Hubble time,~$\Delta\tau = H^{-1}$, thus,~\eqref{kap crit 1} implies~$\kappa_{\rm crit}^{\rm proper} \sim H^3/M_{\rm Pl}^2$. This corresponds to a critical proper lifetime of
\be
\tau_{\rm crit} \sim \frac{M_{\rm Pl}^2}{H^3}\,,
\ee
which is recognized as the Page time for dS space.

Thus, as claimed, the static susceptibility of our set of driven coupled oscillators exhibits a divergence as the average proper lifetime of vacua approaches
the Page time. Translating back to the original volume variables~${\cal V}_i$ suggests an alternative interpretation for the optimal compromise offered by the Page lifetime. Regions
of dS vacua with such lifetimes are most susceptible to (comoving) volume influx from their environment, while at the same time minimizing their MFPT.

\section{Emergence of~$1/f$ Noise}
\label{1/f}

A signature of a self-organized critical system is a~$1/f$ power spectrum for its fluctuations. In this Section we will show that the displacement fluctuations~$q$ for the oscillators in the region
exhibit~$1/f$ signal precisely at the critical Page distribution~\eqref{rho crit}. 

The origin of~$1/f$ signal can be understood intuitively from the linearity of the Ornstein-Uhlenbeck stochastic equation~\eqref{osc 2}. We will see that the power spectrum of~$q$ fluctuations is given by a superposition of Ornstein-Uhlenbeck spectra, weighted by the probability distribution~$\rho(\hat{\kappa})$. For a uniform distribution~\eqref{rho crit}, corresponding to the critical case, we will find that the resulting spectrum is~$1/f$. The derivation is analogous to an old argument to explain the origin of flicker noise as an aggregation of shot noise processes~\cite{vanderziegle}, reviewed in Appendix~D of~\cite{SOCbook}. See also~\cite{origin1overfMarkov,origin1overf}.

\subsection{Auto-correlation function}

Consider the stochastic equation~\eqref{osc 2} governing the oscillator displacements~$q_i$. The solution is trivially found:
\be
q(t) = {\rm e}^{-Kt} q(0) + \int_0^t{\rm d}t'\,{\rm e}^{-K(t-t')} \xi(t')\,.
\label{q(t) soln}
\ee
Since~$\langle \xi(t) \rangle = 0$, the average displacement satisfies
\be
\langle q(t) \rangle = {\rm e}^{-K t} q(0) \xrightarrow[t \rightarrow\infty]{} \; 0\,,
\ee
which follows since~$K$ is positive definite. As expected, the oscillators return to their non-equilibrium steady state position~\eqref{osc eq} as~$t\rightarrow\infty$. 

A more interesting quantity is the unequal time auto-correlation function,
\be
G_{ij}(t) = \lim_{t_0\rightarrow \infty} \langle q_i(t_0) q_j(t+t_0)\rangle\,.
\ee
The large~$t_0$ limit is taken to ensure that~$q$ is statistically stationary, whereas the time difference~$t$ remains fixed. 
Substituting~\eqref{q(t) soln}, the two-point function is
\be
\langle q (t_0) q (t+t_0)\rangle = {\rm e}^{-K t_0} q(0) {\rm e}^{-K (t+t_0)} q(0) + \int_0^{t_0}{\rm d}t' \int_0^{t+t_0}{\rm d}t'' \left\langle {\rm e}^{-K (t_0-t'))} \xi(t') {\rm e}^{-K (t+t_0-t'')} \xi(t'')\right\rangle\,.
\ee
The first term on the right-hand side is negligible in the large~$t_0$ limit and will henceforth be ignored. Substituting the noise correlation~\eqref{noise}, the second term simplifies to give
\be
\langle q(t_0) q(t+t_0)\rangle = \frac{2D}{\Delta t} {\rm e}^{-Kt}  \int_0^{t_0}{\rm d}t' {\rm e}^{-2K(t_0-t')} = D\,  \frac{{\rm e}^{-Kt}}{K\Delta t} \left(1-{\rm e}^{-2Kt_0}\right)  \,.
\ee
In the large~$t_0$ limit we obtain the desired result:
\be
G_{ij} (t) = D\,  \left(\frac{{\rm e}^{-Kt}}{K\Delta t}\right)_{ij}  \,.
\ee
This is independent of~$t_0$, consistent with statistical stationarity. In particular, setting $t = 0$ gives the equal-time two-point function:
\be
G_{ij} (0) = \frac{D}{\Delta t} K^{-1}_{ij}\,.
\label{2pt stat}
\ee

\subsection{Power spectrum}

To proceed it is convenient to average the auto-correlation function by taking the trace:
\be
G(t) \equiv \frac{1}{N} {\rm Tr}\,G(t)  = \frac{D}{N} \sum_\ell  \frac{{\rm e}^{\lambda_\ell t}}{|\lambda_\ell|\Delta t}\,.
\ee
Thus,~$G(t)$ is a linear supposition of exponentially-decaying signals with different characteristic times. The power spectrum is obtained as usual by a cosine transform:
\be
P(\omega) \equiv 2\int_0^\infty {\rm d}t\, G(t) \cos(\omega t) = \frac{2D}{N\Delta t} \sum_\ell \frac{1}{\omega^2 + \lambda_\ell^2}\,.
\ee
Correspondingly~$P(\omega)$ is a superposition of power spectra for Ornstein-Uhlenbeck processes.

Following the steps of Sec.~\ref{chi}, the result greatly simplifies in the downward approximation and continuum limit:
\be
P(\omega) \simeq 2D\,\Delta t  \int_{\hat{\kappa}_{\rm min}}^{\hat{\kappa}_{\rm max}} {\rm d}\hat{\kappa}\, \frac{\rho(\hat{\kappa})}{\hat{\omega}^2 + \hat{\kappa}^2} \,,
\ee
where, as before,~$\hat{\omega} = \omega \Delta t$ and~$\hat{\kappa} = \kappa\Delta t$ are dimensionless variables. This describes a linear supposition of spectra
weighted by the probability distribution~$\rho(\hat{\kappa})$. Focusing on the case where~$\rho(\hat{\kappa})$ is nearly uniform over the relevant range of~$\hat{\kappa}$,
corresponding to the critical distribution~\eqref{rho crit}, the integral can be readily performed:
\be
P_{\rm crit} (\omega) \simeq \frac{2D}{\omega \Delta t} \Big[ \tan^{-1} \left(\frac{\kappa_{\rm max}}{\omega}\right) - \tan^{-1} \left(\frac{\kappa_{\rm min}}{\omega}\right) \Big]\,.
\ee
In the intermediate frequency range~$\kappa_{\rm min} \ll \omega \ll \kappa_{\rm max}$, we have~$\tan^{-1} \frac{\kappa_{\rm max}}{\omega}\simeq \frac{\pi}{2}$ and~$\tan^{-1} \frac{\kappa_{\rm min}}{\omega}\simeq 0$, resulting in
\be
P_{\rm crit} (\omega) \simeq \frac{\pi D}{\omega \Delta t} =  \frac{D}{2\Delta t f}\,,
\ee
where we have translated to ordinary frequency~$f = \frac{\omega}{2\pi}$. Thus, as claimed earlier, fluctuations have a~$1/f$ spectrum at criticality. 

\section{Conclusions}

In this work we have shed light on the nature of the non-equilibrium phase transition at the Page lifetime uncovered in~\cite{Khoury:2019yoo,Khoury:2019ajl}.
We focused on a finite region of the landscape, which exchanges volume with the rest of the landscape and as such acts as an open system. Through
a change of variables we mapped the master equation governing the comoving volume of dS vacua to a stochastic equation for~$N$ coupled overdamped
stochastic oscillators --- the well-known Ornstein-Uhlenbeck process. This allowed us to understand criticality at the Page lifetime as a dynamical phase
transition in the Ornstein-Uhlenbeck process.

We first showed that the static susceptibility of the oscillators diverges as the average of lifetime of dS vacua approaches the Page time. 
A divergent susceptibility is a well-known signature of criticality. In terms of the original volume variables, this corresponds to a divergent
volume susceptibility. In other words, optimal regions of the landscape are most susceptible to volume influx from their environing landscape.

We then showed that the displacement fluctuations for the oscillators exhibit a~$1/f$ power spectrum (or so-called ``pink" noise) over a broad range of frequencies,
precisely at the critical Page distribution. A~$1/f$ power spectrum is another hallmark of non-equilibrium systems at criticality. In analogy with sand
avalanches in the abelian sandpile or neuronal avalanches in the brain, dS vacua at criticality undergo scale invariant volume fluctuation
avalanches.  

The derivation relied on the assumption of a non-vanishing driving term for all sites ($\eta_i\neq 0$ for all $i$) with uncorrelated, white noise fluctuations (though not necessarily Gaussian). It would be interesting to study to what extent this assumption is realistic for optimal regions of the landscape~\cite{Khoury:2019yoo}, {\it i.e.}, with funnel 
topography and Page lifetime distribution. It would also be worthwhile to study other forms of environmental driving, such as non-vanishing driving
for only a subset of sites. More generally, the map to the Ornstein-Uhlenbeck process allows us to cast any approach to the landscape, such as global or local measures,
and, thus, may shed new light on the measure problem.

Our setup satisfies what are believed to be necessary conditions for self-organized criticality: 1)~The region of interest is open and dissipative, in
that dS vacua lose volume to terminals as well as the environment; 2)~The region is maintained out of equilibrium by the environment; and 3) The
environmental driving is very slow compared to the characteristic time scales for transitions between dS vacua in the region. 

Complex self-organized systems poised at criticality are ubiquitous in the natural world~\cite{living}. This has led to the conjecture that
dynamical criticality is evolutionarily favored because it offers an ideal trade-off between robust response to external stimuli
and flexibility of adaptation to a changing environment. Similarly the dynamical mechanism introduced in~\cite{Khoury:2019yoo} selects
regions of the landscape that are tuned at criticality.  

In a forthcoming paper~\cite{to appear} we will study another advantage of dynamical criticality, namely enhanced computational capabilities. 
Indeed, it has been argued that complex systems maximize their computational capabilities at the phase transition between stable
and unstable dynamical behavior --- the so-called ``edge of chaos"~\cite{dynamical crit review}. For instance, cellular automata with
certain critical dynamical rules~\cite{wolfram} are capable of universal computation, exhibiting long-lived and complex transient structures~\cite{edge of chaos}.
In machine learning, a class of recurrent neural networks~\cite{echo state} have been shown to achieve maximal computational power for vanishing
Lyapunov exponent~\cite{RNNchaos}. 

Our region of dS vacua can be thought of as an input-driven dynamical system, with the driving term given by the volume
influx from the environment. As such the region is performing computation. It will be interesting to see whether various information-theoretic
measures of computational efficiency, such as active information storage~\cite{AIS} and information transfer entropy~\cite{transfer entropy}, are maximized at
the critical Page lifetime distribution. This might establish a tantalizing and potentially deep connection between minimal computational complexity of the vacuum
search process~\cite{Denef:2006ad,Carifio:2017bov,Carifio:2017nyb,Halverson:2018cio,Cole:2019enn,Halverson:2020opj} and maximal computational capabilities of optimal landscape regions as input-driven systems.

\vspace{.4cm}
\noindent
{\bf Acknowledgements:} We thank James Halverson, Cody Long, Onkar Parrikar and Sam Wong for enlightening discussions, as well as an anonymous referee for helpful feedback. This work is supported by the US Department of Energy (HEP) Award DE-SC0013528, NASA ATP grant 80NSSC18K0694, and by the Simons Foundation Origins of the Universe Initiative.

\section*{Appendix: Evolution of the Driving Term}

The purpose of this Appendix is to carefully justify the assumptions made for the source term~$\eta$ in the Ornstein-Uhlenbeck equation~\eqref{osc eta}. In particular,
we wish to check under what conditions the evolution of the region can justifiably be treated as Markovian. In what follows we denote the region of interest as the system~S,
and the environment as~E.

By combining~\eqref{master 2} and an analogous equation for~${\cal V}_\alpha$, it is straightforward to obtain the evolution equation
\bea
\nonumber
\frac{{\rm d}\eta_i}{{\rm d}t} &=& \sum_{\alpha,\beta} \kappa_{i\alpha}\left(\kappa_{\alpha\beta} - \delta_{\alpha\beta} \kappa_\beta\right) {\cal V}_\beta \\
 && + \sum_\alpha \kappa_{i\alpha} \sum_j \kappa_{\alpha j} {\cal V}_j\,.
\label{eta eqn}
\eea
In the first line, the total decay rate~$\kappa_\alpha$ includes decays into other dS vacua in~E, to dS vacua in~S, as well as to terminals (in both~S and~E).
Thus the first line describes transitions within~E, as well as decays to vacua inside and outside~E. Crucially, this only depends on the environmental state~${\cal V}_\beta$.  

The second line, on the other hand, describes transitions from~S to some dS vacuum~$\alpha$ in~E, followed by a transition from that same
vacuum $\alpha$ back to~S. Importantly this depends on the state of the system~${\cal V}_j$ and captures how~S perturbs~E.

In order for the evolution of~S to remain Markovian, it is crucial that such backreaction of~S onto~E is negligible. Note that this cannot be justified on the basis of time scales.
The first line only involves transition rates for~E~$\rightarrow$~E/S processes, which are assumed slow; whereas the second line involves~S~$\rightarrow$~E transitions,
which are comparatively faster. 

Instead, the justification relies on the downward approximation. As described in the main text, we imagine that~E consists of two distinct sets of vacua: high energy dS vacua, which source the system, and much lower energy vacua (including terminals), which act as sinks. In this case, the second line in~\eqref{eta eqn} can be neglected because, to leading order in the downward approximation, no vacuum in~E plays a dual role of receiving volume from some vacuum~$j$ in~S and giving it back to another vacuum~$i$ in~S. Neglecting the second line of~\eqref{eta eqn}, the evolution of~$\eta$ becomes independent of the state of~S. Furthermore, since all transition rates in the first line are small,~$\eta$ can be approximated as nearly constant, with small time-dependent fluctuations, as assumed in~\eqref{eta fluc}. 

This situation is akin to other non-equilibrium critical systems. In the idealized example of the abelian sandpile~\cite{SOC1}, sand grains are added to the sandpile, at a rate much
slower than the characteristic time scale for avalanches, and sand grains are also removed through the open boundary at the bottom of the pile during avalanches. Thus~E plays two distinct roles of
adding grains from the top and removing grains from the bottom of the pile. This is quite unlike the standard picture of a system coupled to a thermal bath. In that case,~S-E interactions go both ways, and fast equilibration in~E is necessary to maintain the Markovian nature of~S. In contrast, on the relevant time scale~S and~E are in our case coupled through effectively one-way interactions, thanks to the asymmetry between downward vs upward transitions.

\end{document}